\documentclass{ws-procs9x6}

\begin{document}

\newcommand{\ds}{\displaystyle}
\newcommand{\mc}{\multicolumn}
\newcommand{\bce}{\begin{center}}
\newcommand{\ece}{\end{center}}
\newcommand{\beq}{\begin{equation}}
\newcommand{\eeq}{\end{equation}}
\newcommand{\bea}{\begin{eqnarray}}

\newcommand{\eea}{\end{eqnarray}}
\newcommand{\cont}{\nonumber\eea\bea}
\newcommand{\cl}[1]{\begin{center} {#1} \end{center}}
\newcommand{\ba}{\begin{array}}
\newcommand{\ea}{\end{array}}

\newcommand{\ab}{{\alpha\beta}}
\newcommand{\cd}{{\gamma\delta}}
\newcommand{\dc}{{\delta\gamma}}
\newcommand{\ac}{{\alpha\gamma}}
\newcommand{\bd}{{\beta\delta}}
\newcommand{\abc}{{\alpha\beta\gamma}}
\newcommand{\eps}{{\epsilon}}
\newcommand{\lam}{{\lambda}}
\newcommand{\mn}{{\mu\nu}}
\newcommand{\mpnp}{{\mu'\nu'}}
\newcommand{\Amuu}{{A_{\mu}}}
\newcommand{\Amuo}{{A^{\mu}}}
\newcommand{\Vmuu}{{V_{\mu}}}
\newcommand{\Vmuo}{{V^{\mu}}}
\newcommand{\Anuu}{{A_{\nu}}}
\newcommand{\Anuo}{{A^{\nu}}}
\newcommand{\Vnuu}{{V_{\nu}}}
\newcommand{\Vnuo}{{V^{\nu}}}
\newcommand{\Fmnu}{{F_{\mu\nu}}}
\newcommand{\Fmno}{{F^{\mu\nu}}}

\newcommand{\abcd}{{\alpha\beta\gamma\delta}}


\newcommand{\bsigma}{\mbox{\boldmath $\sigma$}}
\newcommand{\btau}{\mbox{\boldmath $\tau$}}
\newcommand{\brho}{\mbox{\boldmath $\rho$}}
\newcommand{\bpipi}{\mbox{\boldmath $\pi\pi$}}
\newcommand{\bss}{\bsigma\!\cdot\!\bsigma}
\newcommand{\btt}{\btau\!\cdot\!\btau}
\newcommand{\bnabla}{\mbox{\boldmath $\nabla$}}
\newcommand{\bphi}{\mbox{\boldmath $\tau$}}
\newcommand{\bvarphi}{\mbox{\boldmath $\rho$}}
\newcommand{\bDelta}{\mbox{\boldmath $\Delta$}}
\newcommand{\bpsi}{\mbox{\boldmath $\psi$}}
\newcommand{\bPsi}{\mbox{\boldmath $\Psi$}}
\newcommand{\bPhi}{\mbox{\boldmath $\Phi$}}
\newcommand{\bnab}{\mbox{\boldmath $\nabla$}}
\newcommand{\bpi}{\mbox{\boldmath $\pi$}}
\newcommand{\btheta}{\mbox{\boldmath $\theta$}}
\newcommand{\bkappa}{\mbox{\boldmath $\kappa$}}

\newcommand{\bA}{{\bf A}}
\newcommand{\bB}{\mbox{\boldmath $B$}}
\newcommand{\bC}{\mbox{\boldmath $C$}}
\newcommand{\bp}{\mbox{\boldmath $p$}}
\newcommand{\bk}{\mbox{\boldmath $k$}}
\newcommand{\bq}{\mbox{\boldmath $q$}}
\newcommand{\bfe}{{\bf e}}
\newcommand{\bb}{\mbox{\boldmath $b$}}
\newcommand{\bc}{\mbox{\boldmath $c$}}
\newcommand{\br}{\mbox{\boldmath $r$}}
\newcommand{\bR}{\mbox{\boldmath $R$}}

\newcommand{\bs}{{\bf s}}
\newcommand{\bT}{{\bf T}}
\newcommand{\fph}{${\cal F}$}
\newcommand{\aph}{${\cal A}$}
\newcommand{\dph}{${\cal D}$}
\newcommand{\fpi}{f_\pi}
\newcommand{\mpi}{m_\pi}
\newcommand{\Tr}{{\mbox{\rm Tr}}}
\def\Qb{\overline{Q}}
\newcommand{\delu}{\partial_{\mu}}
\newcommand{\delo}{\partial^{\mu}}
%
%
\newcommand{\up}{\!\uparrow}
\newcommand{\upup}{\uparrow\uparrow}
\newcommand{\updo}{\uparrow\downarrow}
\newcommand{\uu}{$\uparrow\uparrow$}
\newcommand{\ud}{$\uparrow\downarrow$}
\newcommand{\auu}{$a^{\uparrow\uparrow}$}
\newcommand{\aud}{$a^{\uparrow\downarrow}$}
\newcommand{\pu}{p\!\uparrow}

\newcommand{\qp}{quasiparticle}
\newcommand{\sa}{scattering amplitude}
\newcommand{\ph}{particle-hole}
\newcommand{\qcd}{{\it QCD}}
\newcommand{\integ}{\int\!d}
\newcommand{\ie}{{\sl i.e.~}}
\newcommand{\etal}{{\sl et al.~}}
\newcommand{\etc}{{\sl etc.~}}
\newcommand{\rhs}{{\sl rhs~}}
\newcommand{\lhs}{{\sl lhs~}}
\newcommand{\eg}{{\sl e.g.~}}
\newcommand{\ef}{\epsilon_F}
\newcommand{\sigt}{d^2\sigma/d\Omega dE}
\newcommand{\sige}{{d^2\sigma\over d\Omega dE}}
\newcommand{\rpaeq}{\beq
\left ( \begin{array}{cc}
A&B\\
-B^*&-A^*\end{array}\right )
\left ( \begin{array}{c}
X^{(\kappa})\\Y^{(\kappa)}\end{array}\right )=E_\kappa
\left ( \begin{array}{c}
X^{(\kappa})\\Y^{(\kappa)}\end{array}\right )
\eeq}
\newcommand{\ket}[1]{| {#1} \rangle}
\newcommand{\bra}[1]{\langle {#1} |}
\newcommand{\ave}[1]{\langle {#1} \rangle}
\newcommand{\half}{{1\over 2}}

\newcommand{\singlespace}{
    \renewcommand{\baselinestretch}{1}\large\normalsize}
\newcommand{\doublespace}{
    \renewcommand{\baselinestretch}{1.6}\large\normalsize}
\newcommand{\bftau}{\mbox{\boldmath $\tau$}}
\newcommand{\bfalpha}{\mbox{\boldmath $\alpha$}}
\newcommand{\bfgamma}{\mbox{\boldmath $\gamma$}}
\newcommand{\bfxi}{\mbox{\boldmath $\xi$}}
\newcommand{\bfbeta}{\mbox{\boldmath $\beta$}}
\newcommand{\bfeta}{\mbox{\boldmath $\eta$}}
\newcommand{\bfpi}{\mbox{\boldmath $\pi$}}
\newcommand{\bfphi}{\mbox{\boldmath $\phi$}}
\newcommand{\bfR}{\mbox{\boldmath ${\cal R}$}}
\newcommand{\bfL}{\mbox{\boldmath ${\cal L}$}}
\newcommand{\bfM}{\mbox{\boldmath ${\cal M}$}}
\def\dblint{\mathop{\rlap{\hbox{$\displaystyle\!\int\!\!\!\!\!\int$}}
    \hbox{$\bigcirc$}}}
\def\ut#1{$\underline{\smash{\vphantom{y}\hbox{#1}}}$}

\def\UNITY{{\bf 1\! |}}
\def\Pom{{\bf I\!P}}
\def\lsim{\mathrel{\rlap{\lower4pt\hbox{\hskip1pt$\sim$}}
    \raise1pt\hbox{$<$}}}         
\def\gsim{\mathrel{\rlap{\lower4pt\hbox{\hskip1pt$\sim$}}
    \raise1pt\hbox{$>$}}}         
\def\beq{\begin{equation}}
\def\eeq{\end{equation}}
\def\bea{\begin{eqnarray}}
\def\eea{\end{eqnarray}}

\title{Nonlinear $k_{\perp}$-factorization: a new paradigm \\
for an in-nucleus hard QCD \footnote{\uppercase{T}his work is supported partly
 by grants \uppercase{ DFG 436 RUS 17/101/04} and 
\uppercase{ DFG RUS 17/138/05}.}}

\author{N.~N. Nikolaev }

\address{ Institut f. Kernphysik, Forschungszentrum J\"ulich, D-52425 J\"ulich, Germany
and L.D.Landau Institute for Theoretical Physics, Chernogolovka, Russia \\
E-mail: N.Nikolaev@fz-juelich.de}

\author{W. Sch\"afer}

\address{Institut f. Kernphysik, Forschungszentrum J\"ulich, D-52425 J\"ulich, Germany
E-mail: Wo.Schaefer@fz-juelich.e}  

\author{B.~G. Zakharov }

\address{ L.D.Landau Institute for Theoretical Physics, Chernogolovka, Russia \\
E-mail: B.Zakharov@fz-juelich.de}
        
\author{V.~R. Zoller }

\address{ Institute for Theoretical and Experimental Physics, 117218, Russia \
E-mail: Zoller@hereon.itep.ru}

\maketitle

\abstracts{
We review the origin, and salient features, of the 
breaking of the conventional linear $k_{\perp}$-factorization 
for an in-nucleus hard pQCD processes. A
realization of the nonlinear $k_{\perp}$-factorization which 
emerges instead is shown to depend on color properties of 
the underlying pQCD subprocesses. We discuss the emerging universality 
classes and extend nonlinear $k_{\perp}$-factorization to 
AGK unitarity rules for the
excitation of the target nucleus.
}

\section{Introduction}

 An extension of factorization theorems to nuclear targets is of
burning urgency - the notion of nuclear gluon densities 
can be made meaningful only if they furnish a unified description
of the whole variety of nuclear hard processes. 
A large thickness of a target nucleus introduces a new scale - the 
so-called saturation 
scale $Q_A^2$, - which breaks the 
familiar linear $k_{\perp}$-factorization 
theorems for hard scattering in a nuclear environment. 
Here we review the recent work by the ITEP-J\"ulich-Landau
collaboration in which a new concept of the nonlinear 
$k_{\perp}$-factorization has been introduced and elaborated
\cite{Nonlinear,PionDijet,SingleJet,Nonuniversality,QuarkGluonDijet,GluonGluonDijet}. 
To this end, we recall the 1975 observation 
\cite{NZfusion}, based on the insight from Gribov \cite{GribovPartons}
and Kancheli \cite{Kancheli}, that
the Lorentz contraction of  ultrarelativistic nucleus entails 
a spatial overlap/fusion/screening of partons from nucleons at 
the same impact parameter if
\beq
x \lsim {{ x_A}}={1/R_A m_N},
\label{eq:1.1}
\eeq
where $R_A$ is the radius of a nucleus of mass number $A$.

Within the fusion reinterpretation of multiple gluon 
exchanges, a nuclear glue
will be a nonlinear functional of the free nucleon
glue: the same sea and glue will be shared by many 
nucleons. Specifically, we derived the collective glue of $j$-overlapping 
nucleons. It is a basis for the definition 
of the collective nuclear unintegrated glue $\phi(b,\bkappa)$, 
per unit area in the impact parameter plane, which 
describes the coherent diffractive dijet production off 
nuclei \cite{NSSdijet,Nonlinear}, for instance, in $\pi A$ collisions
or DIS . We found that the so-defined collective 
nuclear glue provides the linear $k_\perp$-factorization
for the nuclear structure function and the leading single-quark
spectrum, but this is rather an exception due to the Abelian
nature of these observables. All other single-jet and dijet
cross sections prove to be highly nonlinear functionals --
quadratures -- of the collective nuclear glue
\cite{PionDijet,SingleJet,Nonuniversality,QuarkGluonDijet,GluonGluonDijet}.   
Any application 
of linear $k_\perp$--factorization to nuclei
is entirely unwarranted. 
Here we review how the concrete form of the nonlinearity depends 
on the relevant pQCD subprocesses and define the universality
classes of  nonlinear  $k_{\perp}$-factorization for production of
hard dijets. We also report the nonlinear $k_\perp$-factorization
solution\cite{WeAGK}
to the Abramovsky-Gribov-Kancheli (AGK) unitarity 
problem\cite{AGK}.

At the heart of our approach is the equivalence between the parton fusion  
description of the shadowing introduced in 1975 \cite{NZfusion} 
and the unitarization of the color dipole--nucleus interaction 
\cite{NZ91}. The major technical problem in the unitarization program 
is the non-Abelian evolution of color dipoles in a
nuclear environment, and we present a closed-form solution 
based on the multiple-scattering theory for color dipoles 
\cite{Nonlinear,NPZcharm}. A very rich pattern of
nonlinear $k_\perp$-factorization relations emerges already for the
lowest order pQCD subprocesses considered here.


\section{The master formula for nuclear dijets}

In the laboratory frame, dijet production in the parton-nucleus
collisions  can be viewed as an excitation $a\to bc$, where
$a=\gamma^*,q,g,~b,c=q,\bar{q}g$. The parton fusion condition 
(\ref{eq:1.1}) amounts to the coherency of excitation over the
whole diameter of the nucleus. Excitation of the perturbative 
$|bc\rangle$ Fock state of the physical projectile $|a\rangle$ by 
one-gluon exchange, $ag\to bc$, leaves the target nucleon debris in the color 
excited state. For nuclear targets one has to deal with 
multiple gluon exchanges which are enhanced by a large target thickness. 

According to the general rules of quantum mechanics, the
two-particle spectrum is the Fourier transform of the two-body
density matrix. The derivation of the master formula 
for the dijet spectrum, based on the technique developed in
\cite{Nonlinear,NPZcharm} is found in \cite{SingleJet}:
\bea
&&{d \sigma (a^* \to b c) \over dz_b d^2\bp_b d^2\bp_c } = 
{1 \over (2 \pi)^4} \int
d^2\bb_b d^2\bb_c d^2\bb'_b
 d^2\bb'_c  \\
&&\times\exp[-i \bp_b
(\bb_b -\bb'_b) - i
\bp_c(\bb_c
-\bb_c')]
\Psi(z_b,\bb_b -
\bb_c) \Psi^*(z_b,\bb'_b-
\bb'_c) \nonumber \\
&&\times
\Bigl\{
S^{(4)}_{\bar{b}\bar{c} c b}(\bb_b',\bb_c',\bb_b,\bb_c) 
+ S^{(2)}_{\bar{a}a}(\bb',\bb)-
S^{(3)}_{\bar{b}\bar{c}a}(\bb,\bb_b',\bb_c')
- S^{(3)}_{\bar{a}bc}(\bb',\bb_b,\bb_c) \Bigr\}\, .\nonumber
\label{eq:2.1}
\eea 
\begin{figure}[!t]
\begin{center}
\includegraphics[width = 4.0cm,height=11cm,angle=270]{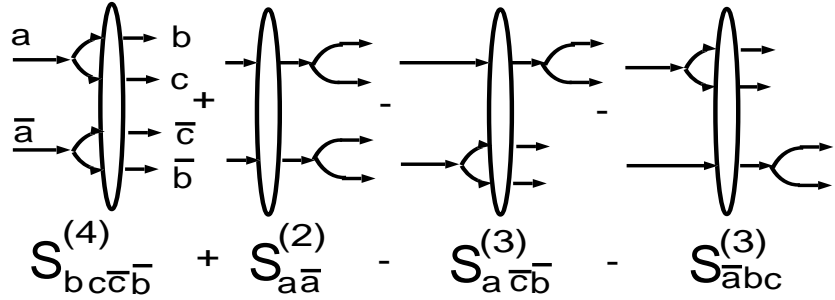}
\caption{The $\textsf{S}$-matrix structure of the two-body density
matrix for excitation $a\to bc$.}
\label{fig:SingleJetDensityMatrix}
\end{center}
\end{figure}
\noindent
If $\bb_a=\bb$ is the projectile's impact parameter, then
$
\bb_{b}=\bb+z_c\br, \quad\bb_{c}=\bb-z_b\br\, ,
$
where $z_{b,c}$ stand for the fraction of the lightcone momentum of the
projectile $a$ carried by partons $b$ and $c$, $\Psi(z,\br)$ stands for
the lightcone wave function of the $\ket{bc}$ Fock state of the projectile,
its connection to the parton-splitting functions is found in \cite{SingleJet}.
All $S^{(n)}$ describe a scattering of color-singlet systems of
$n$ partons, as indicated in Fig.~\ref{fig:SingleJetDensityMatrix}.
This is the crucial point - in the course of our derivation
of the dijet spectra and single-jet spectra we only deal with 
infrared-safe  observables.
$S^{(2)}$
and $S^{(3)}$ are readily calculated in terms of the 2-parton and
3-parton dipole cross sections \cite{NZ91,NZ94,NPZcharm}. 
For the dilute-gas nucleus one uses the Glauber-Gribov multiple-scattering
theory
formula \cite{Glauber,Gribov}
\beq
\textsf{S}^{(n)}({{\bb_c}}',{{\bb_b}}',
{{\bb_c}},{{\bb_b}})=\textsf{S}[\bb,\Sigma_n]=
\exp\{-
{1\over 2}{\Sigma^{(n)}}({{\bb_c}}',{{\bb_b}}',
{{\bb_c}},{{\bb_b}})
T({{\bb}})\},
\label{eq:2.2}
\eeq
where $T(\bb)=\int dr_z n_A(r_z,\bb)$ is the optical 
thickness of a nucleus at the impact parameter $\bb$ and
$n_A(r_z,\bb)$ is the nuclear matter density.
The major nontrivial task is a calculation of the coupled-channel 
operator $\Sigma_{4}$ 
\cite{Nonlinear,QuarkGluonDijet,GluonGluonDijet}. The relevant basis
of color states depends on the pQCD subprocess,
\bea
\gamma^* \to q\bar{q}& :& 
1_1+ 8_{N_c^2},\nonumber\\
 g\to q\bar{q} &:&
1_1+8_{N_c^2}, ,\nonumber\\  q \to qg &:& 3_{N_c} 
+ \{6+15\}_{N_c^3}, \nonumber\\
 g \to gg& : & 1_1+
\{8_A+8_S\}_{N_c^2} + \{10+\overline{10}+27+R_7\}_{N_c^4},\nonumber
\eea
where the subscripts indicate the $N_c$-dependence of the size
of the relevant 
color multiplets. For the description of the two-gluon 
multiplet $R_7$ which exists
for $N_c> 3$ only, see \cite{GluonGluonDijet}. 
These size-dependent subspaces of multiplets prove very useful in
the diagonalization of the non-Abelian evolution of color dipoles.
In the case of nuclear targets
one must distinguish truly inelastic 
processes, which leave the target nucleus in the color excited state,
 and coherent diffraction $aA\to (bc) A$
with retention of the target nucleus in the ground state.


\section{The $k_{\perp}$-factorization for DIS off free nucleons}

The unintegrated gluon density in the
target nucleon, \beq
{\textsf F}(x,\kappa^2) = {\partial G(x, \kappa^2)\over \partial \log \kappa^2},
\label{eq:3.1}
\eeq 
furnishes a universal description of the
proton structure function $F_{2p}(x,Q^2)$ and of the final states. 
For instance, the linear $k_{\perp}$-factorization for forward 
quark-antiquark dijets
reads (for applications, see \cite{NSSSdecor} and references therein)
\bea
&&\frac{2(2\pi)^2d\sigma_N(\gamma^*\to Q\overline{Q})}
{dz d^2\bp
 d^2\bDelta} =   
 f(x, \bDelta )
 \left|\Psi(z,\bp) -
\Psi(z,\bp -
\bDelta )\right|^2 \,,
\label{eq:3.2}
\eea
where 
$\bDelta=\bp_q +
\bp_{\bar{q}}$ is the jet-jet decorrelation (acoplanarity) momentum,
 $\bp\equiv \bp_{\overline{q}},~z\equiv z_{\overline{q}}$ refer to the $\overline{q}$-jet,
and 
\beq
f (x,\bkappa)= {4\pi \alpha_S\over N_c}\cdot {1\over \kappa^4} \cdot {\textsf
F}(x,\bkappa^2)\, . 
\label{eq:3.3} 
\eeq

\begin{figure}[!thb]
\begin{center}
\includegraphics[width = 11.5cm, height= 3.3cm,angle=0]{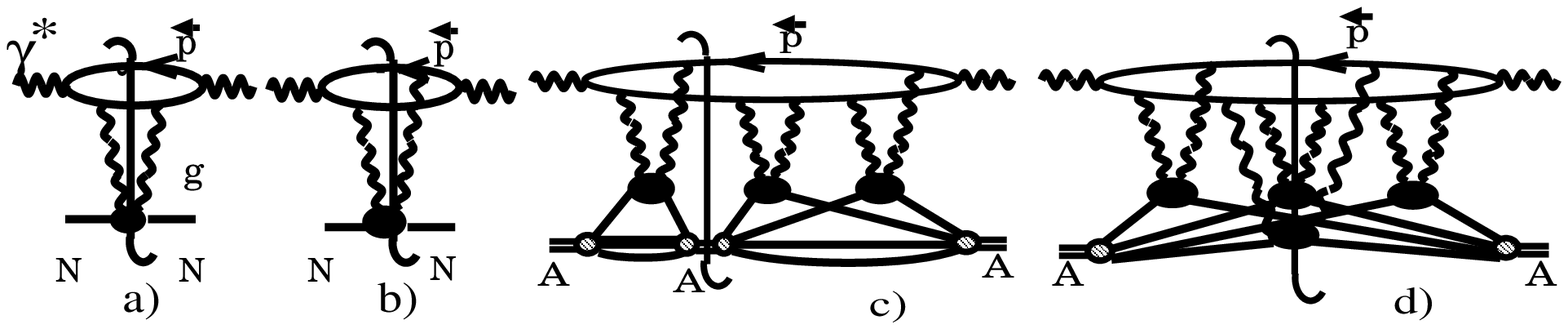}
\caption[*]{ The typical unitarity cuts and dijet final states in DIS : (a),(b) - 
free-nucleon target, (c) - coherent diffractive DIS off a nucleus,
(d) - truly inelastic DIS with multiple color excitation of the nucleus.}
\label{fig:UnitarityDiffraction2004}
\end{center}
\end{figure}

From the unitarity point of view, Eq.~(\ref{eq:3.2}) corresponds to the
unitarity cuts of diagrams Figs. \ref{fig:UnitarityDiffraction2004}a,b 
for the forward Compton scattering amplitude. Notice, that the dijet cross section 
is a direct probe of the free-nucleon unintegrated glue $f (x,\bDelta)$;
the point that the 
jet-jet decorrelation momentum $\bDelta$ comes from the transverse
momentum of the exchanged gluon is obvious form
Figs. \ref{fig:UnitarityDiffraction2004}a,b 
for the forward Compton scattering amplitude. Notice, that the dijet cross section 
is a direct probe of the free-nucleon unintegrated glue $f (x,\bDelta)$;
the point that the 
jet-jet decorrelation momentum $\bDelta$ comes from the transverse
momentum of the exchanged gluon is obvious.
 
The focus of the further discussion is on how linear $k_\perp$-factorization 
(\ref{eq:3.2}) is modified by multiple gluon exchanges
predominant in hard production off nuclear targets.

\section{What is the collective unintegrated nuclear glue?}

While seeking the factorization properties of an in-nucleus hard QCD,
one must define the collective glue in terms of a certain observable
such that the so-defined glue enters in a universal manner a description
of all other nuclear processes. Taking DIS as an example, the same
nuclear glue must describe the total DIS cross section, 
its coherent diffractive unitarity cuts,
Fig. \ref{fig:UnitarityDiffraction2004}c, and multiple color 
excitation unitarity cuts, Fig. \ref{fig:UnitarityDiffraction2004}d,
of the nuclear forward Compton scattering amplitude. 

DIS off a nucleus at $x\lsim x_A$ can be described in terms
of the color dipole-nucleus cross-section 
\cite{NZ91} 
\beq
{{\sigma_A(\br)}} = 2\!\!\int\!\!
d^2{{\bb} }\Big\{1 -\exp[-\frac{1}
{2}
{{\sigma(x,\br)}}
T({{\bb}})]\Big\}\,,
\label{eq:4.1}
\eeq
where 
\beq
\sigma(x,\br) = \int d^2\bkappa f (x,\bkappa)
[1-\exp(i\bkappa\br)] 
\label{eq:4.2}
\eeq 
is the $q\bar{q}$
dipole cross section \cite{NZ91}.  The $x$-dependence of 
$\sigma(x,\br)$ is governed by the color-dipole evolution
equation \cite{NZ94,NZZBFKL}; for the sake of brevity, here-below we
suppress the $x$-dependence. The nuclear dipole
cross section (\ref{eq:4.1}) sums in the compact form the Glauber-Gribov
multiple-scattering diagrams and is a basis for the
quantitative description of nuclear shadowing in DIS \cite{BaroneShad}.

It is remarkable ,that although a deposition of dozen
MeV energy will break any heavy nucleus, at $x\lsim x_A$ 
such a coherent
diffraction makes $\approx 50 \%$ of the total cross section
of DIS off heavy nucleus \cite{NZZdiffr}. To the lowest
order in pQCD, the coherent diffractive final state consists
of the back-to-back dijet with vanishing transverse momentum
transfer $\bDelta$ to the target nucleus and the large transverse 
momentum $\bp$ of dijets in $\pi A$ collisions 
comes entirely from gluons exchanged with target
nucleons \cite{NSSdijet,NZsplit}. Consequently, one can take the
partial wave of the diffraction amplitude, i.e., the nuclear
profile function, for the definition of  the collective 
nuclear glue per
{{ unit area}} in the impact parameter space,
${{\phi}}({{\bb}},{{\bkappa}}
) $
\cite{NSSdijet,Nonlinear}:
\bea 
\Gamma_A({{\bb}},{{\br}})&=&[1 -\exp(-\frac{1}
{2}
{{\sigma(\br)}}T({{\bb}}))]
=\int d^2{{\bkappa}}
{{\phi}}({{\bb}},{{\bkappa}}
) \{1-\exp[i {{\bkappa}}
{{\br}} ]\} \,. 
\label{eq:4.3} 
\eea
In can be expanded as \beq
{{\phi}}({{\bb}},{{\bkappa}}
) = {1\over \sigma_0}\sum_{j=1}^{\infty}{{
w_{j}({{\bb}})}}
 f^{(j)}({{\bkappa}}),
\label{eq:4.4}
\eeq
where the collective glue $f^{(j)}({{\bkappa}} )$
of $j$ overlapping nucleons of the Lorentz-contracted nucleus satisfies the 
convolution representation 
\beq
f^{(j)}(\bkappa) ={1\over \sigma_0} (f^{(j-1)} \otimes f)(\bkappa), 
\label{eq:4.5}
\eeq
and
a probability
to find  $j$ overlapping nucleons equals
\beq
w_{j}(\bb) ={
\nu_{A}^j({{\bb}})
 \over j!} \exp\left[-\nu_{A}({{\bb}})\right],~~\nu_{A}({{\bb}}) =  
{1\over 2}\sigma_0 {{T(\bb)}},
\label{eq:4.6}
\eeq
 $\sigma_0 =\int d^2\bkappa f(\bkappa)$ is the dipole cross section $\sigma(\br)$
for large dipoles $\br$.

Subsequently we  will use
also $
{{\Phi}}({{\bb}},{{\bkappa}}
) = {{\phi}}({{\bb}},{{\bkappa}}
)+ w_0(\bb)\delta({{\bkappa}} )$, which is the Fourier
transform of the $\textsf{S}$-matrix for an elementary
dipole,

\bea
\textsf{S}[\bb,\sigma(\br)]
=\int d^2{{\bkappa}}
{{\Phi}}({{\bb}},{{\bkappa}}) \exp[i {{\bkappa}}
{{\br}} ] \,. 
\label{eq:4.7} 
\eea
The antishadowing properties \cite{NSSdijet,Nonlinear} of
the so-defined collective glue  $\phi(\bb,\bkappa)$ 
are responsible for the familiar Cronin effect \cite{SingleJet}.
It furnishes 
the linear $k_{\perp}$-factorization for the nuclear structure function
$F_{2A}(x,Q^2)$ and the forward 
single jets in DIS off nuclei, precisely the same as for the 
free-nucleon target in terms of $f(\bkappa)$. This linear $k_{\perp}$-factorization 
is rather an exception 
because of their special Abelian features.
Dijets in DIS, and dijet and single-jet spectra in $hA$
collisions, however, prove to be 
highly nonlinear functionals of $\Phi({{\bb}},{{\bkappa}})$.
\cite{Nonlinear,PionDijet,Nonuniversality,SingleJet,QuarkGluonDijet,GluonGluonDijet}.  
We dubbed this property the nonlinear $k_\perp$-factorization.

An important point is that color dipole cross section depends 
on the $SU(N_c)$ representation the partons belong to. For this
reason the collective nuclear glue can not be described by 
one scalar function, it is rather a density matrix in the space
of color representations, for examples see \cite{SingleJet,GluonGluonDijet}.


\section{The origin of nonlinear $k_\perp$-factorization}

 The fundamental point is that the calculation
of the single-jet and dijet cross sections can be reduced to an
interaction with the nucleus of color-singlet multiparton systems
\cite{NPZcharm}, as exemplified by our master formula
(\ref{eq:2.1}). The interaction cross-section for such systems of 
$n$ partons is a (matrix) superposition of elementary dipole 
cross-sections. Upon the diagonalization of the non-Abelian
evolution problem, the matrix elements of nuclear $\textsf{S}$-matrices
$\textsf{S}^{(n)}$, $n\geq 3$, in (\ref{eq:2.1}) will be a (matrix) products 
of nuclear $\textsf{S}$-matrices  for elementary dipoles. 
In the calculation of the Fourier transform of the two-body density
matrix,  nuclear $\textsf{S}$-matrices for elementary dipoles
will be replaced by their Fourier transforms (\ref{eq:4.7}). 
Consequently, the Fourier transforms of $\textsf{S}^{(n)}$ would
give rise to multiple convolutions of $\Phi(\bb,\bkappa)$, and the
nuclear dijet spectra will be highly nonlinear quadratures of 
collective nuclear glue. The specific expansion for eigenvalues
of $\textsf{S}^{(n)}$ in term of elementary dipole cross sections,
and the specific form of the emerging nonlinear $k_\perp$-factorization 
thereof, depends on the
color representations of the initial-state parton and final-state 
dijet, still there emerges a well-defined pattern of the 
universality classes.


\section{Why collective glue for slices of a nucleus is a must?}

In the realm of pQCD with partons in the fundamental and adjoint
multiplets, one finds an exact product representations for nuclear  
$\textsf{S}$-matrices for $n=2,3$ in terms of the $\textsf{S}$-matrices
for elementary dipoles. The case of $n=4$ is much more tricky. It
is remarkable how the forbidding three-channel for $qg$ dijets, and
seven-channel for $gg$ dijets, non-Abelian intranuclear evolution
can be diagonalized exactly for large $N_c$ due  the block-diagonal
structure of the four-parton cross-section operator $\hat{\Sigma}^{(4)}$. 
We explicitly
represent it in the form 
\bea \hat{\Sigma}^{(4)} = \hat{\Sigma}^{(0)} + \hat{\omega} \, ,
\label{eq:6.1}
\eea
where $\hat{\Sigma}^{(0)}$ is the diagonal operator, 
while the off-diagonal $\hat{\omega}$, proportional to
the transition cross section $\Omega(\bs,\br,\br')$
introduced in \cite{Nonlinear,QuarkGluonDijet},  describes  
the $1/N_c$ suppressed  transitions between the subspaces
formed by the states  having equal dimension 
at large $N_c$ as expounded in Sec. 2.
The expansion (\ref{eq:6.1}) is a a basis for the systematic 
large-$N_c$ perturbation theory. One can address it, for instance,
making use of the Sylvester expansion \cite{Nonlinear}, which has
a generic form (we show the result for the 
case of $gg$-dijets, the detailed description of eigenstates $\ket{e_i}$
and eigenvalues $\Sigma_i$ is found in \cite{GluonGluonDijet})
\bea
&&\textsf{S}[\bb, \hat{\sigma}^{(4)}(\bs,\br,\br')]= \ket{e_2}\bra{e_2} 
\exp[-{1\over 2} \Sigma_2 T(\bb)] \\
\label{eq:6.2}
&&+ \ket{e_4}\bra{e_2} {\Omega(\bs,\br,\br') \over \sqrt{2}N_c(\Sigma_2-\Sigma_4)}
\left\{\exp[-{1\over 2} \Sigma_4 T(\bb)]-\exp[-{1\over 2} \Sigma_2 T(\bb)]\right\}. 
\nonumber
\eea
Here $\Sigma_{2,4}$ -- the relevant eigenvalues of $\hat{\Sigma}^{(0)}$ --
are certain superpositions of the elementary dipole cross sections,
While the Fourier transform of the diagonal term in (\ref{eq:6.2}) will
be a multiple convolution of the collective nuclear glue, for the 
off-diagonal transitions, which excite $gg$-dijets in higher color
multiplets, $10, \overline{10},27,R_7$, this is the dead end, though.
The denominator $(\Sigma_2-\Sigma_4)$ in the 
Sylvester expansion (\ref{eq:6.2}) blocks further analytic
calculations. For instance, in their closely related study,
Kovchegov and Jalilian-Marian \cite{Kovchegov}
stop at precisely this point; in order 
to proceed further one has to rely upon the 
brute force numerical Fourier transform, which masks important distinction 
between different production mechanisms and salient features of the
production dynamics.

The way to the explicit analytic quadratures is paved by our
integral representation \cite{Nonlinear}, which 
to the first order in the off-diagonal perturbation $\hat{\omega}$
takes the form (similar representation is found for higher orders
of large-$N_c$ perturbation theory \cite{Nonlinear})
\bea \textsf{S}[\bb, \hat{\Sigma}^{(0)} + \hat{\omega}]
-\textsf{S}[\bb, \hat{\Sigma}^{(0)}] = -{1 \over 2} T(\bb) \int_0^1
d\beta \, \textsf{S}[\bb, (1- \beta)  \hat{\Sigma}^{(0)}] \,
\hat{\omega} \, \textsf{S}[\bb, \beta \hat{\Sigma}^{(0)}]. \nonumber\\
\label{eq:6.3}
\eea
This integral representation is much more than a technical trick to
achieve the convolution representation for the four-body $\textsf{S}$-matrix.
Specifically, $\hat{\omega}$ describes the hard excitation of dijets
in higher color multiplets  at the depth $\beta$ from the front
face of the nucleus, while $ \textsf{S}[\bb, \beta \hat{\Sigma}^{(0)}]$
and $\textsf{S}[\bb, (1- \beta)  \hat{\Sigma}^{(0)}]$ describe the Initial 
(ISI) and Final State Interaction (FSI) in slices $[0,\beta]$ and 
$[\beta,1]$ of the nucleus, respectively. Now, the
color multiplet of  the dijet is different from that of the incident 
parton, and the resulting distinction between ISI and FSI is an integral part
of the dijet production dynamics, which can not be described entirely 
in terms of the classical gluon field of a whole nucleus \cite{CGC}. 
The emergence of $ \textsf{S}[\bb, \beta \hat{\Sigma}^{(0)}]$
and $\textsf{S}[\bb, (1- \beta)  \hat{\Sigma}^{(0)}]$ calls upon  
the nuclear glue defined for the slice 
$[0,\beta]$ of the nucleus ( $0\leq \beta \leq 1$)
\beq
\exp\left[-{1\over 2}{{\beta}}
\sigma(\br)T(b)\right]=\int d^2{{\bkappa}} 
{{\Phi({{\beta}};\bb,{{\bkappa}})}}
\exp(i{{\bkappa}}\br),
\label{eq:6.4}
\eeq 
and the corresponding overlap probabilities $w_j(\beta;\bb)$. 
We will need also the wave function of $bc$ Fock states coherently 
distorted in this slice, 
\beq
\Psi({{\beta}};z,\bp)\equiv
\int {{d^2\bkappa}} {
{\Phi({{\beta}};\bb,{{\bkappa}})}}
\Psi(z,\bp +\bkappa).
\label{eq:6.5}
\eeq 
Now we are in the position to present  our explicit results on
nonlinear $k_\perp$-factorization. We start with the short digression
into the single-jet problem.


\section{The fate of $k_{\perp}$-factorization for single-jet spectra
in $pA$ collisions}

The integration over the transverse momentum of the unobserved parton
in the master formula (\ref{eq:2.1}) is straightforward. The
unobserved parton and its antiparton will enter at the same impact 
parameter and multiparton color singlet states simplify to the
two-parton ones.  Still, the non-Abelian features of QCD manifest
themselves in the breaking of linear $k_{\perp}$-factorization.
Here we show the the large-$N_c$ result for the radiation of gluons 
from quarks, $q^*\to qg$ \cite{SingleJet}, which is directly relevant 
to jet production in the proton hemisphere of $pA$ collisions at RHIC
\cite{STAR,BRAHMS}. The linear $k_\perp$-factorization holds for the free-nucleon
target, 
\bea 
&& \frac{(2\pi)^2 d\sigma_A(q^* \to g q) }{ dz_g d^2\bp_g }=
 \frac{1}{ 2}\int d^2\bkappa f(\bkappa) \\
&&\times \Big\{ \bigl|
\Psi(z_g,\bp_g) - \Psi(z_g,\bp_g+\bkappa)\bigr|^2 + \bigl|\Psi(z_g,\bp_g+\bkappa) -
\Psi(z_g,\bp_g+z_g \bkappa)\bigr|^2 \Big\} \, ,\nonumber
\label{eq:7.1}
\eea
while the same spectrum for the nuclear target is of the two-component form
\bea &&\frac{ (2\pi)^2 d\sigma_A(q^* \to g q) }{ dz_g d^2\bp_g
d^2\bb} = S[\bb,\sigma_0]
\int d^2\bkappa \phi(\bb,\bkappa)
\\\label{eq:7.2}
&&\times \Big\{|\Psi(z_g,\bp_g) - \Psi(z_g,\bp_g+\bkappa) |^2
+ |\Psi(z_g,\bp_g+\bkappa) - \Psi(z_g,\bp_g+z_g \bkappa)|^2\Big\} +\nonumber\\
&& \int d^2\bkappa_1 d^2\bkappa_2 \phi(\bb,\bkappa_1) \phi(\bb,\bkappa_2)
|\Psi(z_g,\bp_g+z_g \bkappa_1) - \Psi(z_g,\bp_g+\bkappa_1 + \bkappa_2)|^2 \, .
\nonumber
\eea
The first component is an exact counterpart of the free-nucleon spectrum:
it is linear $k_{\perp }$-factorizable, but is suppressed by the nuclear 
absorption factor $S[\bb,\sigma_0(x)]$. For central interactions of the 
main experimental interest, the gluon spectrum is entirely dominated by the second 
component which is a non-linear -- quadratic -- functional of the collective nuclear 
glue. A full compendium of nonlinear $k_{\perp}$-factorization
results for single jets from all possible pQCD subprocesses is
found in \cite{SingleJet}.

There is a conspicuous difference
between the  $z_g$-dependence of the free-nucleon and nuclear spectra.
This amounts to the $\bp_g$-dependence of the Landau-Pomeranchuk-Migdal
effect; the same applies to the spectrum of leading quarks and
nuclear quenching of forward jets in $pA$ collisions \cite{SingleJet}.


\section{Nonlinear $k_\perp$-factorization for dijets: the case of DIS}

Here we report closed-form analytic results for quark-antiquark dijets
in DIS to the leading order in $1/N_c$,
higher orders  can be derived following Ref. \cite{Nonlinear}.
In DIS off nuclei,  dipoles propagate as color-singlets until 
at depth $\beta$ from the front face of a nucleus
they excite into the octet state; the non-Abelian evolution
in the slice $[\beta,1]$ consists of color rotations 
within the octet state. Combining the truly inelastic and
diffractive components of DIS,
\bea
&&\frac{(2\pi)^2d\sigma_{A}(\gamma^*\to Q\overline{Q}) }{ d^2\bb dz d^2\bp d^2\bDelta} = \frac{1}{
2} T(\bb) 
\int_0^1 d \beta
\int d^2\bkappa_1 d^2\bkappa 
\nonumber\\
&&\times f(\bkappa)\Phi(1-\beta,\bb,\bDelta -\bkappa_1 -\bkappa)
\Phi(1-\beta,\bb,\bkappa_1)\nonumber\\
&&\times \Bigl|
\Psi(\beta;z,\bp -\bkappa_1) -
\Psi(\beta; z,\bp  -\bkappa_1-\bkappa)
\Bigr|^2
\nonumber\\
&&+ \delta^{(2)}(\bDelta)\Bigl|
\Psi(1;z,\bp) -
\Psi( z,\bp)\Bigr|^2\, .
\label{eq:8.1}
\eea
The diffractive component of Eq. (\ref{eq:8.1}), $\propto \delta^{(2)}(\bDelta)$, gives exactly 
back-to-back dijets (for 
 the $\bDelta$-dependence for finite-size nuclei see  
Ref. \cite{NSSdijet}). It is a quadratic functional of the
collective nuclear glue. The first component in
(\ref{eq:8.1}) -- truly inelastic DIS with excitation
of color-octet dijets -- is of fifth order in gluon field densities: 
a linear one of the free-nucleon glue $f(\bkappa)$
which describes the hard singlet-to-octet transition and
a quartic functional of  the collective nuclear glue for the two slices of a 
nucleus: the coherent ISI's of color-singlet dipoles in
the slice $[0,\beta]$ and incoherent FSI's of color-octet dipoles in the slice $[\beta,1]$.


\section{Nonlinear $k_\perp$-factorization for dijets: universality classes}

We start with the universality class of coherent diffraction.
The origin of coherent diffractive DIS is a difference between the in-vacuum
and intranuclear-distorted wave function of the $q\bar{q}$ state
of the photon:
\bea
\frac{(2\pi)^2d\sigma_{A}(\gamma^*\to q\overline{q}) }{ d^2b dz d^2\bp d^2\bDelta} =  
{\delta^{(2)}(\bDelta)}\Bigl|
{ \Psi(1;z,\bp)} -
\Psi( z,\bp)\Bigr|^2\, .
\label{eq:9.1}
\eea
In the case of $q\to qg$ the incident partons are colored, 
and intranuclear attenuation of the incident quark wave 
entails the suppression of coherent diffraction by the factor
$w_0^2(b)$, similar, but stronger intranuclear attenuation is found for
$g\to gg$ \cite{GluonGluonDijet}. 
This nuclear suppression of coherent diffraction can be 
identified with Bjorken's diffractive gap survival probability \cite{Bjorken},
its process dependence makes clear that the concept of 
hard factorization for the nuclear pomeron makes no sense.  
In all above cases coherent diffraction is
of leading order in $1/N_c$, while coherent diffraction  $g\to q\bar{q}$ 
is suppressed, which adds further case against the hard factorization
for nuclear pomerons.

The universality class of dijets in higher color miltiplets excited 
from partons in lower multiplet is the most interesting 
one. As an example we cite the large-$N_c$ result \cite{QuarkGluonDijet}
\bea
&&{d \sigma \bigl(q^* \to qg(6+15)\bigr) 
\over d^2b dz d^2\bDelta d^2\bp }
 = {1 \over 2(2 \pi)^2} T(b)\int_0^1 d{\beta}  \nonumber\\
&\times& \int  d^2\bkappa d^2\bkappa_1 d^2\bkappa_2 d^2\bkappa_3
\delta(\bkappa +\bkappa_1+\bkappa_2+\bkappa_3-\bDelta)\nonumber\\
&\times& {{\underbrace{\Phi({\beta};\bb,\bkappa_3)}_{Quark~~ ISI}}}
 \underbrace{ f(\bkappa)\left|\Psi({\beta};z,\bp-\bkappa_2-\bkappa_3)-
\Psi({\beta};z,\bp-\bkappa_2-\bkappa_3-\bkappa)\right|^2}
_{Hard ~~Excitation}
\nonumber\\
&\times&{{\underbrace{\Phi(1-{\beta};\bb,\bkappa_1)}_{Quark~~ FSI}}}
{{\underbrace{\Phi({C_A\over C_F}(1-{\beta});\bb,\bkappa_2)}_{Gluon~~ FSI}}},
\label{eq:9.2}
\eea 
where we indicated the r\^ole of different factors in the
integrand. Notice the sixth order nonlinearity of the dijet 
spectrum (\ref{eq:9.2}) in nuclear and free-nucleon glue. 
The most nontrivial point here is how incoherent distortions 
of the incident quark wave come along with the coherent
distortions of the $qg$ wave function in the same slice $[0,\beta]$
of the nucleus.

The ratio 
of Casimirs $C_A/C_F$ in 
$\Phi({C_A\over C_F}(1-\beta),\bb,\bkappa_2)$
for FSI of gluons 
is a reminder that the collective nuclear glue
derives from a density matrix in  
color space \cite{Nonlinear,SingleJet}.
In DIS, $\gamma^* \to  q\bar{q}(8)$, ISI of the photon
vanishes and $\Phi({\beta};\bb,\bkappa_3)=\delta^{(2)}(\bkappa_3)$,
for the incident gluons, $g \to  gg(10+\overline{10}+27+R_7)$, 
the quark FSI factor must be swapped for the gluon FSI factor, and
there will be a slight modification of incoherent ISI distortions 
- at large $N_c$ the gluon behaves like the uncorrelated 
quark-antiquark pair. 
\cite{SingleJet,GluonGluonDijet}. The striking distinction between ISI and FSI 
requires the collective nuclear glue for slices of the nucleus --
nonlinear $k_\perp$-factorization can not be described by 
the classical gluon field of the whole nucleus.
 The nonlinear $k_{\perp}$-factorization 
entails a nuclear enhancement of the decorrelation 
of dijets \cite{Nonlinear,QuarkGluonDijet,GluonGluonDijet}.

Universality class of dijets in the same lower color multiplet
as the beam parton has its own unique features.
For instance, in the case of color-triplet $qg$ dijets \cite{QuarkGluonDijet}
\bea
{d \sigma \bigl(q^*A \to qg(3)\bigr) \over d^2b dz d^2\bDelta d^2\bp }
& =& {1 \over (2 \pi)^2}\phi(b,\bDelta)
\left| { \Psi(1;z,\bp-\bDelta)}-  { \Psi(z,\bp-z\bDelta)}\right|^2.\nonumber\\
\label{eq:9.3}
\eea 
The free-nucleon cross section has exactly the same form 
in terms of $f(\bDelta)$
and the 
in-vacuum wave functions. Recalling that  { $\Psi(z,\bp-z\bDelta)$},
which has a collinear singularity, 
is the probability amplitude to find
the $qg$ state in the physical quark, we reinterpret this result as
a fragmentation of the quark, scattered quasi-elastically 
with the differential cross section $\propto \phi(b,\bDelta)$,
see also below. The change 
from the free-nucleon to a nuclear target, 
$$|\Psi(z,\bp-\bDelta)- 
\Psi(z,\bp-z\bDelta)|^2 \Longrightarrow
| \Psi(1;z,\bp-\bDelta)- 
\Psi(z,\bp-z\bDelta)|^2,$$ 
must be interpreted as a nuclear modification
of the hard fragmentation function of the quark.

Apart from the fact that ISI of incident gluons looks 
like an ISI of the uncorrelated 
quark-antiquark pair, the nonlinear $k_\perp$-factorization 
for $g \to q\bar{q}(8)$, $g\to gg(8_A+8_S)$,
$g\to gg(8_S)$ is very similar to that of $q\to qg(3)$ \cite{GluonGluonDijet}.
In $g\to gg(8_A+8_S)$ too there emerges the nuclear modified
hard fragmentation function of the gluon.


\section{Nonlinear $k_\perp$-factorization for dijets:
unitarity cuts and AGK rules for color excitation of the target nucleus}

Our technique can readily be extended \cite{WeAGK} to partial cross
sections $d\sigma_{j}$ for final states with $j$ 
color-excited nucleons of the target nucleus -- in the language of the so-called 
AGK unitarity rules 
that corresponds to $j$ cut pomerons \cite{AGK}. This multiplicity $j$ 
controls the hadronic multiproduction in the nucleus 
hemisphere, i.e., the collision centrality, as well as the 
nonperturbative energy-loss 
contribution to the quenching of forward
jets. The simplest case is the AGK rule for the universality class of 
coherent diffraction:
\beq
d\sigma_{j} = \delta_{ j 0}d\sigma_D.
\label{eq:10.00}
\eeq

Now we notice that the differential cross section of 
quark-nucleon quasi-elastic scattering $q N\to q' {N^*}$ equals 
\beq
{d\sigma_{qN}\over d^2\bkappa}= {1\over 2} f(\bkappa)
.
\label{eq:10.0}
\eeq
 Here the target 
debris ${N^*}$ is in the color-excited state. Then, the 
convolution property (\ref{eq:4.5}) of the collective 
glue
$f^{(j)}(\bkappa)$ relates it to the differential
cross section of $j$-fold incoherent quasi-elastic scattering,
\beq
f^{(j)}(\bkappa) \propto {d\sigma^{(j)}\over d^2\bkappa}.
\label{eq:10.1}
\eeq
Evidently, the latter is simply a $j$-fold convolution of
single scattering cross sections. 
This observation paves a way to  
a simple interpretation of $f^{(j)}(\bkappa)$ in
terms of $j$ cut pomerons.

We illustrate the emerging AGK rule for the universality class of 
dijets in the same lower color representation as the beam parton 
on an example of the reaction $q\to qg(3)$ with $j$ color-excited
nucleons of the target nucleus:
\bea
&&{d \sigma_{j} \bigl(q^*A \to qg(3) \bigr) 
\over d^2b dz d^2\bDelta d^2\bp }=\nonumber\\
&&=
{1 \over (2 \pi)^2\sigma_0} 
 w_{j}(b)f^{(j)}(\bDelta)
\left| { \Psi(1;z,\bp-\bDelta)}-  { \Psi(z,\bp-z\bDelta)}\right|^2.
\label{eq:10.2} 
\eea 
It makes the interpretation of this final state as a result 
of fragmentation of the quasi-elastically scattered quark an
obvious one - the nuclear-distorted fragmentation function,
its collinear pole included,
does not depend on the multiplicity of cut pomerons $j$. 
Notice the uncut pomerons which enter the 
intranuclear distortion of $\Psi(1;z,\bp-\bDelta)$.

As an example of AGK rules from the universality class of dijets in higher 
color multiplet excited from partons in lower multiplet we
show the large-$N_c$ result for $q\to qg(6+15)$ with $j$ color-excited nucleons
of the target nucleus:
\bea
&&{d \sigma_j \bigl(q^* \to qg(6+15)\bigr) 
\over d^2b dz d^2\bDelta d^2\bp }
 = {1 \over (2 \pi)^2} T(b)\int_0^1 d{\beta}  \nonumber\\
&\times& \int  d^2\bkappa d^2\bkappa_1 d^2\bkappa_2 d^2\bkappa_3
\delta(\bkappa +\bkappa_1+\bkappa_2+\bkappa_3-\bDelta)\nonumber\\
&\times& 
\sum_{n,k,m} \delta(j-n-k-m-1){\underbrace{ w_{m}(\beta;\bb) 
{f^{(m)}(\bkappa_3)\over \sigma_0}}_{Quark ~~ISI}
}\nonumber\\
&\times& \underbrace{ f(\bkappa)\left|\Psi({\beta};z,\bp-\bkappa_3-\bkappa_2)-
\Psi({\beta};z,\bp-\bkappa_3-\bkappa_2-\bkappa)\right|^2}_{Hard~~ excitation}\nonumber\\
&\times&
{\underbrace{w_k \Big({C_A\over C_F}(1 -{\beta});\bb\Big) 
{f^{(k)}(\bkappa_2)\over \sigma_0}}
_{Gluon~~ FSI}}\times
{\underbrace{w_{n} \Big((1 -{\beta});\bb\Big)
 {f^{(n)}(\bkappa_1)\over \sigma_0}}_{Quark~~ FSI}}
\label{10.3} 
\eea 
It corresponds to the following counting of cut pomerons: one
 cut pomeron for hard excitation at the depth ${\beta}$;
 $m$  cut pomerons for $m$-fold quasi-elastic 
intranuclear scattering 
of the incident quark in the slice $[0,{\beta}]$;
 $k$  cut pomerons for $k$-fold quasi-elastic 
intranuclear scattering 
of the final-state gluon in the slice $[{\beta},1]$
and  $n$  cut pomerons for $n$-fold quasi-elastic 
intranuclear scattering 
of the final-state quark in the slice $[{\beta},1]$ of the nucleus.
There are more uncut pomerons which describe the coherent distortion
of the $qg$ wave function in the slice $[0,\beta]$. The
functional form of the hard excitation factor, which can also
be re-interpreted as the (non-local) gluon radiation vertex,
is preserved. However, upon the convolution with
partial collective glue for initial-state and final-state slices
of the nucleus, its numerical value will depend on
the multiplicity of cut pomerons in the intimal-state and
final-state channels.

New features of our AGK rules for an in-nucleus hard QCD are noteworthy:
(i) we established simple relationship between cut pomerons, 
quasi-elastic scattering of partons and collective glue for
overlapping nucleons of the Lorentz-contracted nucleus,
(ii) the concrete realization of the AGK rules depends on the
universality class the specific pQCD subprocess belongs to,
(iii) the coherent nuclear distortions described by uncut
pomerons persist in all universality classes, (iv) the
gluon radiation vertex depends on the multiplicity of cut
pomerons.

\section*{Conclusions}

 Hard processes in a nuclear environment must be
described by nonlinear $k_\perp$-factorization in terms of the
collective nuclear glue defined through the coherent diffractive
dijet production. 
Any application of the familiar linear $k_\perp$-factorization 
would be entirely erroneous. We reported explicit quadratures for single-jet 
to dijet spectra from all pQCD subprocesses.
Indispensable virtue of nonlinear 
$k_\perp$-factorization is  a nontrivial interplay of incoherent
rescatterings and coherent distortions of the
dijet wave function. Still another important virtue is that it requires
collective glue for slices of a nucleus - 
nonlinear $k_\perp$-factorization can not be described by
a classical gluon field of the whole nucleus. Our 
connection between the collective glue for overlapping nucleons and
the cross section of multiple quasi-elastic scattering
of partons gives a simple interpretation of the AGK unitarity
rules for excitation of the target nucleus. These unitarity rules can be applied to
evaluation of the energy loss and quenching of forward jets.

\section*{Acknowledgments}
This presentation is is a humble tribute to the memory of 
a great theorist Vladimir N.~Gribov, aka BH, whose technique 
we used heavily in the reported work. We hope BH would have liked the results.
NNN is grateful to J.~Nyri, Yu.~Dokshitzer and P.~Levai 
for the invitation to this Workshop.

\end{document}